\documentclass[preprint,showpacs,secnumarabic,amssymb,superscriptaddress,prl,aps]{revtex4-1}
\usepackage{float}
\usepackage{amsmath}
\usepackage{graphicx}
\usepackage{epstopdf}
\usepackage{dcolumn}
\usepackage{bm}
\usepackage{color}
\usepackage{mathtools}
\usepackage{braket}
\usepackage{url}
\usepackage[colorlinks,linkcolor=blue,anchorcolor=blue,citecolor=blue]{hyperref}

\begin{document}

\title{Intense harmonics with time-varying orbital angular momentum from relativistic plasma mirrors}
\author{Jingwei Wang}
\email{wangjw@siom.ac.cn}
\affiliation{State Key Laboratory of High Field Laser Physics and CAS Center for Excellence in Ultra-intense Laser Science, Shanghai Institute of Optics and Fine Mechanics(SIOM), Chinese Academy of Sciences(CAS), Shanghai 201800, China}
\affiliation{Collaborative Innovation Center of IFSA, Shanghai Jiao Tong University, Shanghai 200240, China}
\author{Matt Zepf}
\affiliation{Helmholtz Institute Jena, Fr\"{o}belstieg 3, 07743 Jena, Germany}
\affiliation{Faculty of Physics and Astronomy, Friedrich-Schiller-Universit\"{a}t Jena, 07743 Jena, Germany}
\author{Yuxin Leng}
\author{Ruxin Li}
\affiliation{State Key Laboratory of High Field Laser Physics and CAS Center for Excellence in Ultra-intense Laser Science, Shanghai Institute of Optics and Fine Mechanics(SIOM), Chinese Academy of Sciences(CAS), Shanghai 201800, China}

\author{Sergey G. Rykovanov}
\email{S.Rykovanov@skoltech.ru}
\affiliation{High Performance Computing and Big Data Laboratory, Center for Computational and Data-Intensive Science and Engineering, Skolkovo Institute of Science and Technology, Moscow, 121205, Russia}

\begin{abstract}
In this Letter using three-dimensional particle-in-cell simulations and analytical considerations we demonstrate intense high-order plasma surface harmonics carrying a time-varying orbital angular momentum (OAM) -- the self-torque. We show that by using two laser beams with different OAMs $l_1$ and $l_2$ and a certain delay between each other and shooting them obliquely on an overdense plasma target, one can generate harmonics with OAM spanning $nl_1$ to $nl_2$, where $n$ is the order of the harmonic. Such intense self-torqued harmonics can offer new possibilities in ultrafast spectroscopy.
\end{abstract}
\date{\today}
\maketitle
Angular momentum (spin or orbital) is one of the properties of light, along with the intensity, frequency, etc. Historically there were several important milestones in the development of understanding and usage of light angular momentum. In 1909 Poynting showed using the mechanical analogy that a circularly polarized light carries a spin angular momentum (SAM)~\cite{Poynting1909}. In 1992 Allen et al. suggested that a light beam with a helical phase-front carries an orbital angular momentum (OAM)~\cite{Allen1992}. Over the past few decades considerable attention has been given to developing methods to control and manipulate the OAM of light beam, such as imparting OAM onto light~\cite{Biener2002, Sueda2004}, transfer between SAM and OAM~\cite{Marrucci2006, WangNC2019}, and generating extreme-ultraviolet (EUV) high-order harmonics with well-defined OAM~\cite{Hernandez2013, Gariepy2014, Zhang2015, Geneaux2016, Gauthier2017, Denoeud2017, Leblanc2017}. The rapid development of OAM light beams is driven by their important applications in optical manipulation~\cite{Padgett2011}, optical microscopy~\cite{Maurer2010,furhapter2005}, and optical communications~\cite{Wang2012, Willner2015}.

The transverse phase structure of a beam with OAM, also known as a vortex beam, is typically described by a certain dependence of an electromagnetic field component on azimuthal angle $\phi$ proportional to $\exp(-il\phi)$, where integer $l$ is the so-called topological charge number. The OAM of a vortex beam is then characterized by $l\hbar$. For a long time the OAM of light was recognized as a static quantity without time-dependence, until recently L. Rego et al. discovered a new class of OAM light beams: the self-torqued light beams carrying time-varying OAM~\cite{Rego2019}. The self-torqued light beam was produced from the nonlinear process of high-order harmonic generation in gases driven by two time-delayed laser pulses carrying different values of OAM. The self-torque $\hbar\xi$ is defined as $\hbar\xi=\hbar dl(t)/dt$, where $l(t)$ is the time-dependent OAM of the light beam. Therefore, the term self-torque refers to the angular acceleration of the light beam, in an analogy with other physical systems that possess a self-induced time variation of the angular momentum, such as the gravitational self-fields~\cite{Dolan2007}. This inherent property of light opens new routes for creating structured light beams. The self-torqued light beams could be used for investigating the systems with time-varying OAM, such as imaging magnetic and topological excitations, launching selective and chiral excitation of quantum matter, imprinting OAM centrifuges as well as for ultrafast spectroscopy.

\begin{figure*}[htbp]
\begin{center}
\includegraphics[width=0.9\linewidth]{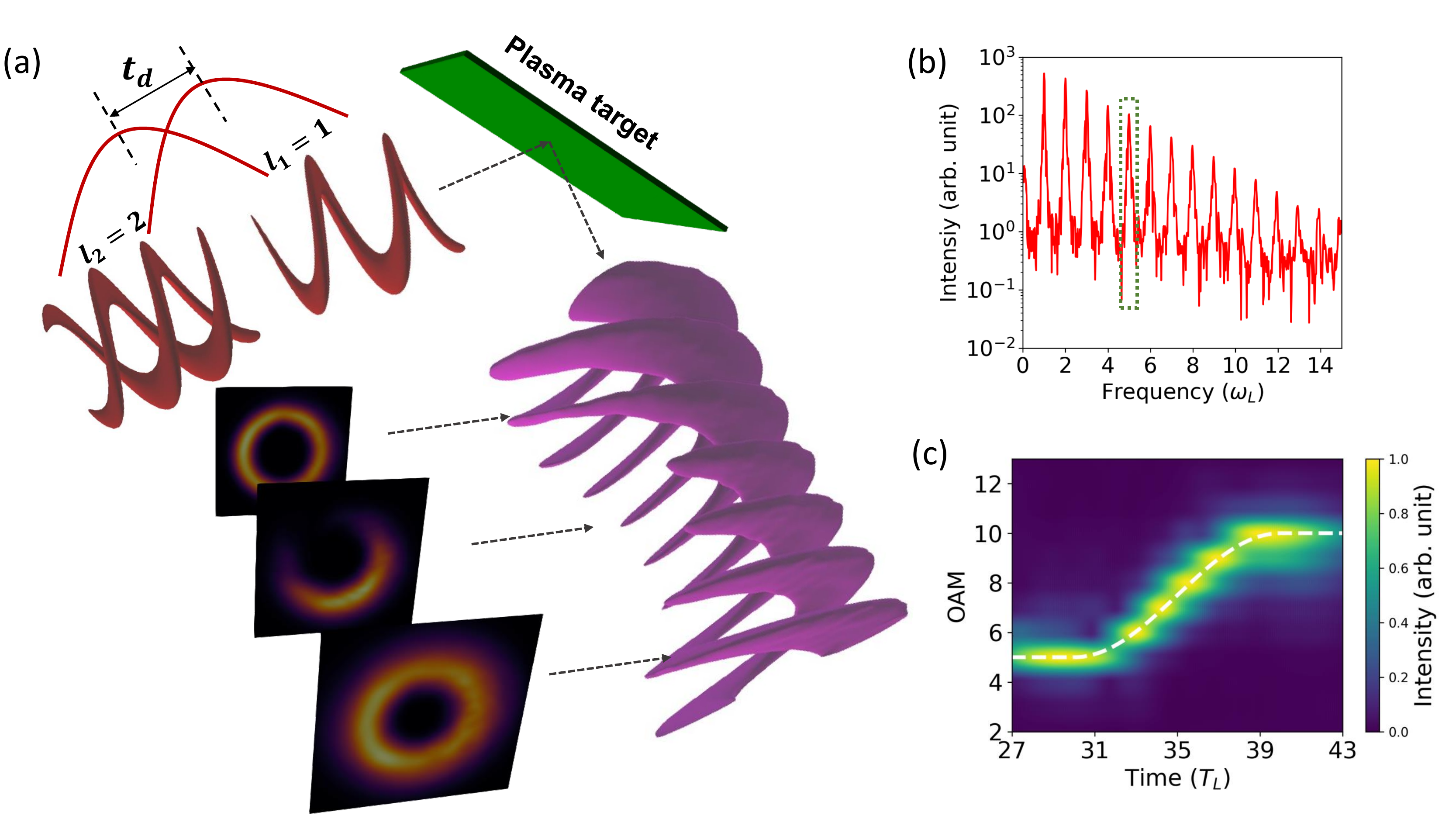}
\centering
\caption{(a) Schematic of generation of self-torqued harmonics from the oscillating plasma mirrors driven by two relativistic lasers with different values of OAM $l_1$ and $l_2$. The embedded three slices are the intensity distributions of the harmonic at different longitudinal positions. (b) The spectrum of the reflected laser field at some fixed radial coordinate. (c) Temporal evolution of the OAM of the 5-th order harmonic. The white dashed line is the theoretical result calculated from Eq.~(\ref{eq:e7}). The parameters of lasers and target are given in the simulation setup in the text.}
\label{fig1}
\end{center}
\end{figure*}

In \cite{Rego2019} the self-torqued EUV harmonics are driven by moderately intense laser pulses (with intensities $\sim 10^{14}~\rm{W/cm}^2$) in gas. The low intensity of the generated self-torqued EUV beams would be a limitation for their future applications. Plasma surface harmonics, on the other hand, could be much more intense as they are driven by relativistic lasers (with intensities more than $10^{18}~\rm{W/cm}^2$)~\cite{Teubner2009}. Generation of intense surface harmonics carrying OAM in the relativistic region has already been experimentally demonstrated~\cite{Denoeud2017, Leblanc2017}. However, it is not clear whether it is possible to generate self-torqued high-order harmonics in the relativistic region, where the involved physics is quite different from the gas harmonics generation. In this Letter, we reveal that self-torqued EUV beams naturally arise when the plasma surface harmonics generation is driven by two relativistic and time-delayed laser pulses with different values of OAM, via the well-known relativistic oscillating mirror (ROM) mechanism~\cite{Bulanov1994, Lichters1996}.

The interaction schematic is demonstrated in Fig.~\ref{fig1}(a): two $p$-polarized, time-delayed intense vortex laser beams carrying different OAM values ($l_1$ and $l_2$) are obliquely incident onto a solid target. The driving laser field oscillates the plasma surface with a velocity close to the light speed. At the same time, the oscillating plasma surface reflects the driving field and modulates it, which in turn brings multiple harmonics into the reflected field, as shown in Fig.~\ref{fig1}(b). In the hand-waiving view of quantum physics, the harmonics generation can be understood as a process of absorbing $q$ photons with a frequency $\omega$ and emitting one photon with a frequency $q\omega$. According to the conversation of angular momentum, the emitted photon will carry an OAM of $ql$, given that each absorbed photon carries an OAM of $l$ ($\hbar$ is dropped here and also in the following text). For the present scheme, there are simultaneously two kinds of photons in the overlapping region of the time-delayed lasers, in terms of the photon OAM values: $l_1$ and $l_2$. Assuming that the emitted photon with a frequency of $q\omega$ comes from $m$ photons with OAM $l_1$ and $(q-m)$ photons with OAM $l_2$, then each photon in the $q$-th harmonic will carry an OAM of $ml_1+(q-m)l_2$. Since $m$ is related with the ratio of the instantaneous intensities of the two pulses, $m$ is time-dependent, and consequently, the OAM of the $q$-th harmonic is also time-dependent. This reasoning gives an intuitive understanding of the generation of time-varying OAM in high order harmonics.

The self-torque can also be explained from the theory of ROM mechanism~\cite{Bulanov1994, Lichters1996, Tsakiris2006}. The oblique incidence in the laboratory frame can firstly be reduced to a normal incidence in a moving frame by the Bourdier transform~\cite{Bourdier1983}. In the moving frame we assume that on the target at some radial coordinate the electric fields of the two $p$-polarized driving lasers are $E_1=u_1(t)\exp[i(\omega_L t-k_L z+l_1\phi)]$ and $E_2=u_2(t)\exp[i(\omega_L t-k_L z+l_2\phi)]$, where $\omega_L$ is the laser circular frequency, $k_L$ is the wave vector and $u_1(t)$ and $u_2(t)$ are the field amplitudes. The total field can be written as
\begin{equation}
E(\phi, t) = E_1 + E_2 = u_0(t)e^{i\omega_L t - k_L z}[(1-\eta)e^{il_1\phi}+\eta e^{il_2\phi}],
\label{eq:e1}
\end{equation}
where we have defined $u_0(t) = u_1(t) + u_2(t)$ and $\eta = u_2(t)/u_0(t)$. For convenience, we factorize $E(\phi, t)$ as
\begin{equation}
E(\phi, t) = |E(\phi,t)| e^{i\omega_L t - k_L z + i\varphi(\phi,t)},
\label{eq:e2}
\end{equation}
with the phase $\varphi(\phi,t) \approx (1-2\eta)(l_1 - l_2)\phi/2 + (l_1 + l_2)\phi/2 = [(1-\eta)l_1 + \eta l_2]\phi$, while the amplitude $|E(\phi,t)|$ will not be discussed in the present problem. According to the ROM theory, the reflected field observed at ($t, z<0$) is emitted at a retarded time $t^{\prime} = t-Z(t^\prime)/c+z/c$ from the oscillating source located at $Z(t^\prime)$, where $Z(t^\prime)$  is the longitudinal trajectory of the plasma mirror. Thus the reflected field at the observer can be written as
\begin{equation}
E_r(\phi, t) \propto |E_(\phi,t)| e^{i(\omega_L t - 2k_L Z(t^\prime)) + i\varphi(\phi,t)}.
\label{eq:e33}
\end{equation}
Here we have dropped the fixed phase $k_Lz$. The plasma mirror driven by the laser field will oscillate as $k_LZ(t^\prime) = \varepsilon [(1-\eta)\sin(\omega_L t^\prime + l_1\phi)) + \eta \sin(\omega_L t^\prime +l_2\phi)]$, where $\varepsilon$ is the amplitude. With the expression of $Z(t^\prime)$ we then rewrite Eq.~(\ref{eq:e33}) as

\begin{equation}
\begin{aligned}
E_r(\phi, t^\prime) \propto & |E(\phi,t^\prime)|e^{i\omega_L t^\prime + i [(1-\eta)l_1+\eta l_2]\phi} 
 e^{-i\varepsilon [(1-\eta)\sin(\omega_L t^\prime + l_1\phi) + \eta\sin(\omega_L t^\prime + l_2\phi)]}.
\end{aligned}
\label{eq:e3}
\end{equation}
We then work in the $t^\prime$ coordinate for convenient. Employing the Jacobi-Anger identity\cite{Cuyt2008}, we can rewrite the last exponential term of Eq.~(\ref{eq:e3}) as
$\sum_{n=-\infty}^{n=+\infty} J_n[\varepsilon(1-\eta)] \exp[in(\omega_L t^\prime + l_1\phi+\pi)] \sum_{m=-\infty}^{m=+\infty} J_m(\varepsilon \eta) \exp[im(\omega_L t^\prime + l_2\phi+\pi)]$, $J_n$ and $J_m$ are the Bessel functions of the first kind. Putting this expression into Eq.~(\ref{eq:e3}), we get the field of the $(q+1)$-th order harmonic as
\begin{equation}
\begin{aligned}
E_r^{q+1} & \propto e^{i(q+1)\omega_L t^\prime} e^{i [(1-\eta)l_1+\eta l_2]\phi} 
\sum_{n=0}^{q}J_n[\varepsilon(1-\eta)]J_{q-n}(\varepsilon \eta) e^{i[nl_1+(q-n)l_2]\phi}.
\end{aligned}
\label{eq:e4}
\end{equation}
From Eq.~(\ref{eq:e4}) we see that each term in the sum over $n$ has a definite OAM of $nl_1+(q-n)l_2$, and each term contributes to the mean OAM with a weight of $J_n[\varepsilon(1-\eta)]J_{q-n}(\varepsilon \eta)$. Then the mean OAM of the $(q+1)$-th order harmonic at time $t^\prime$ can be calculated as
\begin{equation}
\begin{aligned}
\overline{l}_{q+1}(t^\prime) & = [(1-\eta)l_1 + \eta l_2] 
+\frac{\sum_{n=0}^{q} J_n[\varepsilon(1-\eta)] J_{q-n}(\varepsilon \eta) [n l_1 + (q-n) l_2]}{\sum_{n=0}^{q} J_n[\varepsilon(1-\eta)] J_{q-n}(\varepsilon \eta)}.
\end{aligned}
\label{eq:e5}
\end{equation}
Since $\varepsilon <1$, we can use the approximation $J_v(x) \approx(1/v!)(x/2)^v$. Eq.~(\ref{eq:e5}) can be written as
\begin{equation}
\begin{aligned}
\overline{l}_{q+1}(t^\prime) &= [(1-\eta)l_1 + \eta l_2] 
+ \frac{\sum_{n=0}^{q} \binom{q}{n} (1-\eta)^n \eta^{q-n}[nl_1 + (q-n)l_2]}{\sum_{n=0}^{q} \binom{q}{n} (1-\eta)^n \eta^{q-n}}, \\
\end{aligned}
\label{eq:e6}
\end{equation}
where $\binom{q}{n}=q!/[n!(q-n)!]$ is the binomial coefficient. Considering that $\sum_{n=0}^{q} \left ( ^q _n \right )(1-\eta)^n \eta^{q-n} = 1$ and $\sum_{n=0}^{q} \left ( ^q _n \right )(1-\eta)^n \eta^{q-n}n = q(1-\eta)$, we get the final result of the mean OAM for the $(q+1)$-th order harmonic as
\begin{equation}
\overline{l}_{q+1}(t^\prime) = (q+1)[(1-\eta)l_1 + \eta l_2].
\label{eq:e7}
\end{equation}
This result shows that the $(q+1)$-th harmonic contains a time-varying OAM, changing from $(q+1)l_1$ to $(q+1)l_2$ over time.

The self-torque has been verified by three-dimensional particle-in-cell simulations using the code LAPINE~\cite{lapine2002}. Two time-delayed laser pulses carrying OAM $l_1=1$ and $l_2=2$ are obliquely incident on a solid target with an incident angel $\alpha=45^\circ$. The envelopes of the two incident laser pulses are shaped with the same function $\sin^2(\pi t/T_0)~(0\leq t\leq T_0)$, with the pulse full duration $T_0=20 T_L$. $T_L=\lambda_L/c$ is the laser period, $\lambda_L=0.8~\mu {\rm m}$ is the laser wavelength and $c$ is the light speed in the vacuum. The time delay between the two pulses is chosen as $t_d = T_0/2$ for the main results. The normalized amplitudes of the two laser fields $E_L$ are both $a_0 = e|E_L|/m_e\omega_L c = 1.68$, where $m_e$ is the electron mass and $-e$ is the electron charge. The target is plane with a thickness of $2\lambda_L$ and a density of $5n_c$, where $n_c=1.1\times10^{21}\lambda_L(\mu {\rm m})/{\rm cm}^3$ is the critical density. The size of the simulation box is $20\lambda_L(x)\times20\lambda_L(y)\times20\lambda_L(z)$ corresponding to grids $1000(x)\times1000(y)\times1000(z)$, with 8 macro-particles per cell. A detector in the $x-z$ plane is placed at $y=0.5\lambda_L$ to collect the reflected field.

Fig.~\ref{fig1}(b) shows the harmonic spectrum of the reflected field at some fixed radial coordinate. The intensity of each harmonic can also be calculated. For instant, the intensity of the $5$-th order harmonic (with a wavelength of $160\rm~{nm}$) is $1.37\times10^{17}~\rm{W/cm}^2$, which shows the harmonics are indeed intense. We choose the $5$-th order harmonic as an example to check the self-torque. The distribution of its OAM during the overlapping time is presented in Fig.~\ref{fig1}(c). One can see that the harmonic contains all the OAM values between $l=5$ and $l=10$. The most probable value of OAM (i. e. the OAM value which has the peak intensity at a time) increases with time. The time evolution agrees well with the theoretical result from Eq.~(\ref{eq:e7}), shown by the white dashed line in Fig.~\ref{fig1}(c). The intensities of each OAM are calculated from the Fourier analysis of the electric field upon the azimuth angel. An animation describing the temporal evolution of the field profile of the $5$-th order harmonic can be found in the Supplemental Material and it shows how the OAM value of $5$-th harmonic changes from $l=5$ to $l=10$.

\begin{figure}
\begin{center}
\includegraphics[width=0.95\linewidth]{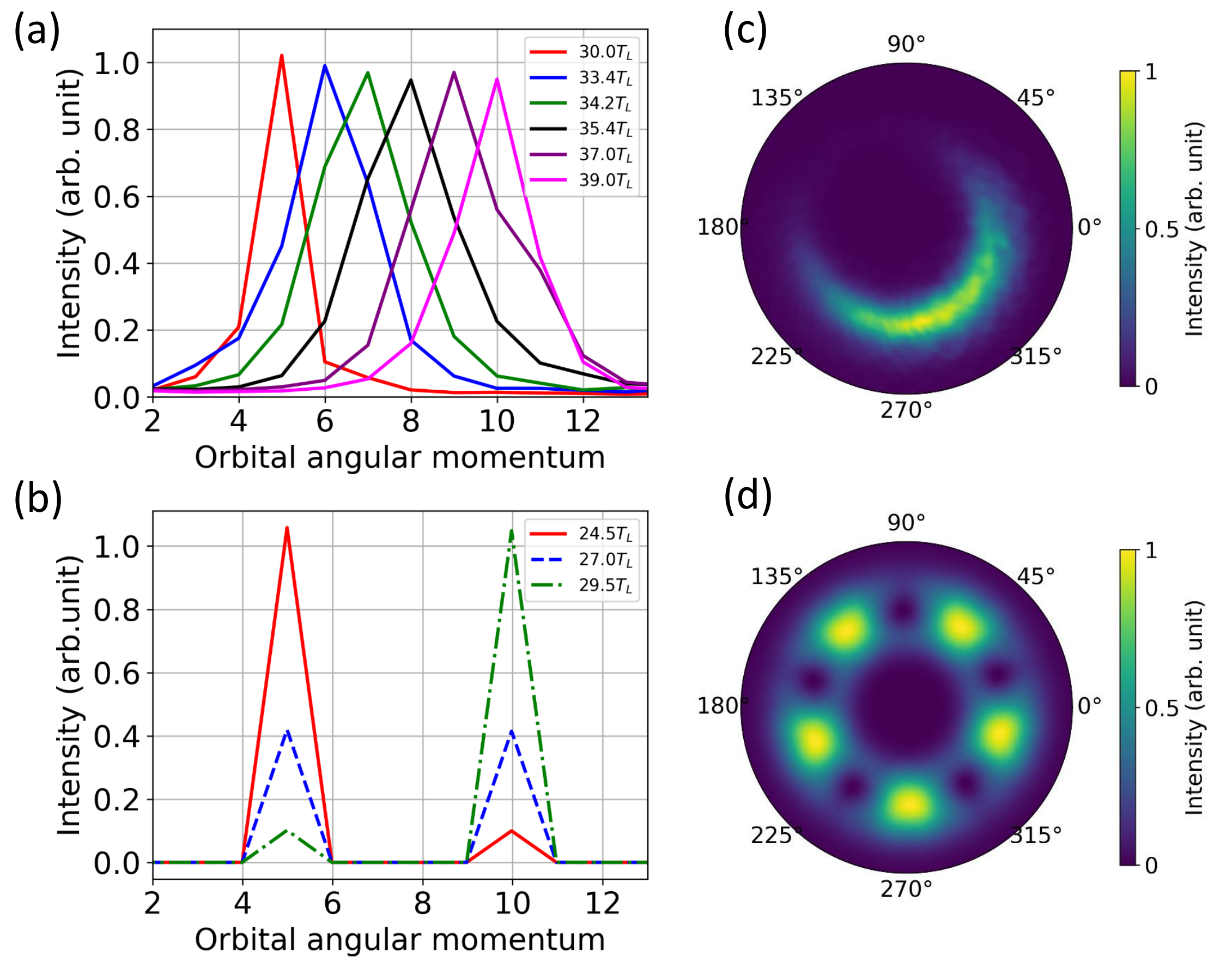}
\centering
\caption{The distinctions between a self-torqued beam and a fractional vortex beam. The OAM spectrums at different times for a self-torqued beam (a) and a fractional vortex beam (b). The intensity distribution of a self-torqued beam (c) and a fractional vortex beam (d). The self-torque beam corresponds to the 5-th harmonic in Fig.~\ref{fig1}. The fractional vortex beam corresponds to two time-delayed vortex beams with $l_1=5$, $l_2=10$, pulse duration $T_0=20T_L$ and time delay $t_d=T_0/2$.}
\label{fig2}
\end{center}
\end{figure}

It is important to emphasize that such a self-torqued beam differs essentially from the so-called fractional OAM beams. A fractional OAM beam, which is a superposition of two time-delayed vortex beams with OAM $ql_1$ and $ql_2$, does not contain a self-torque, although the temporal evolution of its average OAM is similar with that of a self-torqued beam. An important evidence is that, a fractional OAM beam does not contain physical intermediate OAM states (i.e., photons with OAM other than $ql_1$ and $ql_2$), while a self-torqued beam contains all the intermediate OAM states between $ql_1$ and $ql_2$. This is demonstrated by the OAM content at different times in Fig.~\ref{fig2}(a) and (b). In Fig.~\ref{fig2}(a) we plot the OAM spectra for the self-torqued $5$-th order harmonic, while Fig.~\ref{fig2}(b) is for the overlapping of two time-delayed OAM beams with $l=5$ and $l=10$. One can clearly see that there are six OAM values (from $l$=5 to $l$=10) during the overlapping time for a self-torque beam, while there are only two OAM values during the whole interaction time for the fractional OAM beam. The intermediate OAM values of the self-torqued beam come from the interactions of the two driving lasers with different OAMs, which is indicated in Eq.~(\ref{eq:e4}). Another distinction between a self-torqued beam and a fractional beam is the intensity distribution. Fig.~\ref{fig2}(c) and (d) show the intensity distribution of the $5$-th order harmonic of the self-torque beam and a fractional beam with $l=5$ and $l=10$. One can see that the intensity of a self-torqued beam exhibits a distinctive ``crescent'' shape.

\begin{figure}
\begin{center}
\includegraphics[width=1.0\linewidth]{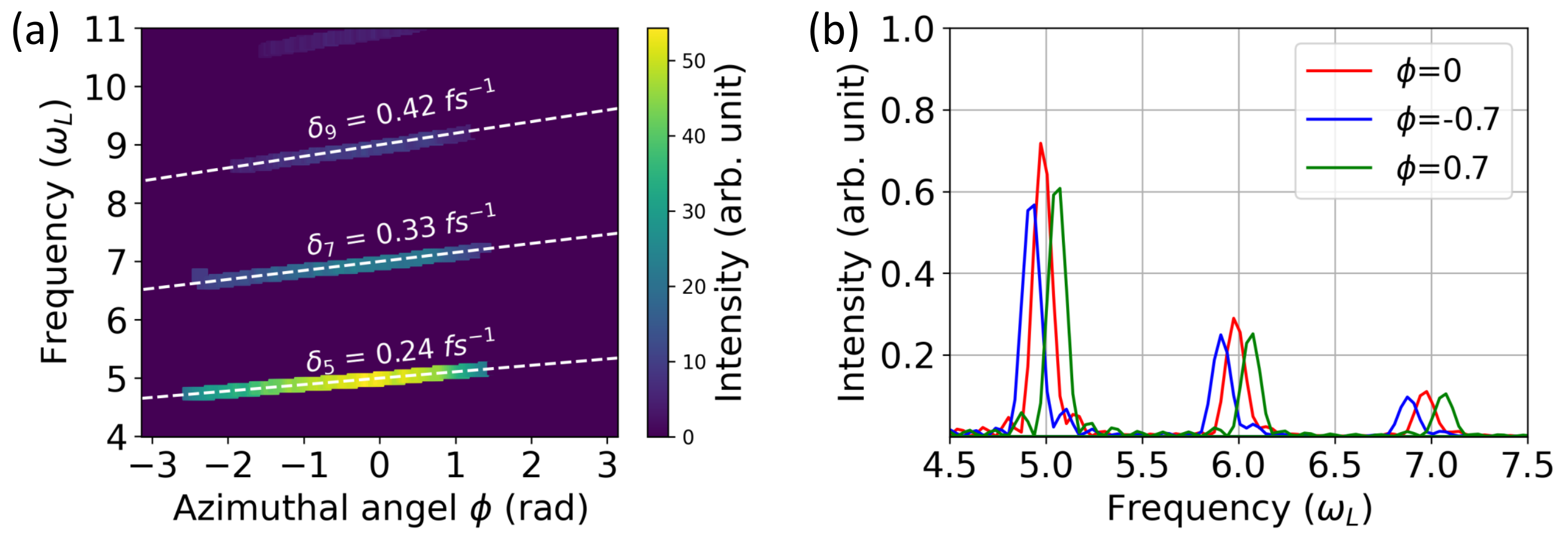}
\centering
\caption{The azimuthal frequency chirp. (a) The distribution of harmonic intensity in the space of frequency and azimuthal angel for the $5$-th,  $7$-th and $9$-th order harmonics. The dependences of frequency on the azimuthal angel from the theoretical model are demonstrated by the white lines. The self-torque here is the average over the overlapping time. (b) The harmonic spectrums at different azimuthal angles. The simulation parameters here are the same as Fig.~\ref{fig1}. }
\label{fig3}
\end{center}
\end{figure}

A consequence of the self-torque is the presence of an azimuthal frequency chirp, which means that the same order harmonic at different azimuthal angles will carry slightly different frequencies. The phase of the $q$-th order harmonic can be written as $\varphi_q = q\omega_L t + l_q\phi$. Since $l_q$ is time-dependent, the frequency of the $q$-th order harmonic at an azimuthal angle $\phi$ will then be shifted as
\begin{equation}
\omega_q(\phi) = \frac{d \varphi_q}{dt}  \approx  q\omega_L + \frac{d\overline{l}_q(t)}{dt}\phi = q\omega_L  + \xi_q \phi.
\label{eq:e8}
\end{equation}
The self-torque here is $\xi_q=d\overline{l}_q(t)/dt$, where $\overline{l}_q(t)$ is the mean OAM calculated by Eq.~(\ref{eq:e7}). Fig.~\ref{fig3}(a) shows the intensity distributions of the harmonics in the space of frequency and azimuthal angle. For each harmonic, the frequency almost linearly increases with the azimuthal angel $\phi$. The simulation results agree well with the theoretical results demonstrated by the white dashed lines in Fig.~\ref{fig3}(a). One can also find that the self-torque is proportional to the harmonic order. The spectra in Fig.~\ref{fig3}(b) show that the frequencies at $\phi=\pm0.7$ are about $0.05\omega_L$ shifted and their intensities are a bit lower, as compared to $\phi=0$. In experiments, the azimuthal frequency chirp can be confirmed by the phenomenon that photons will have different energies at different azimuthal angles.

\begin{figure}
\begin{center}
\includegraphics[width=1.0\linewidth]{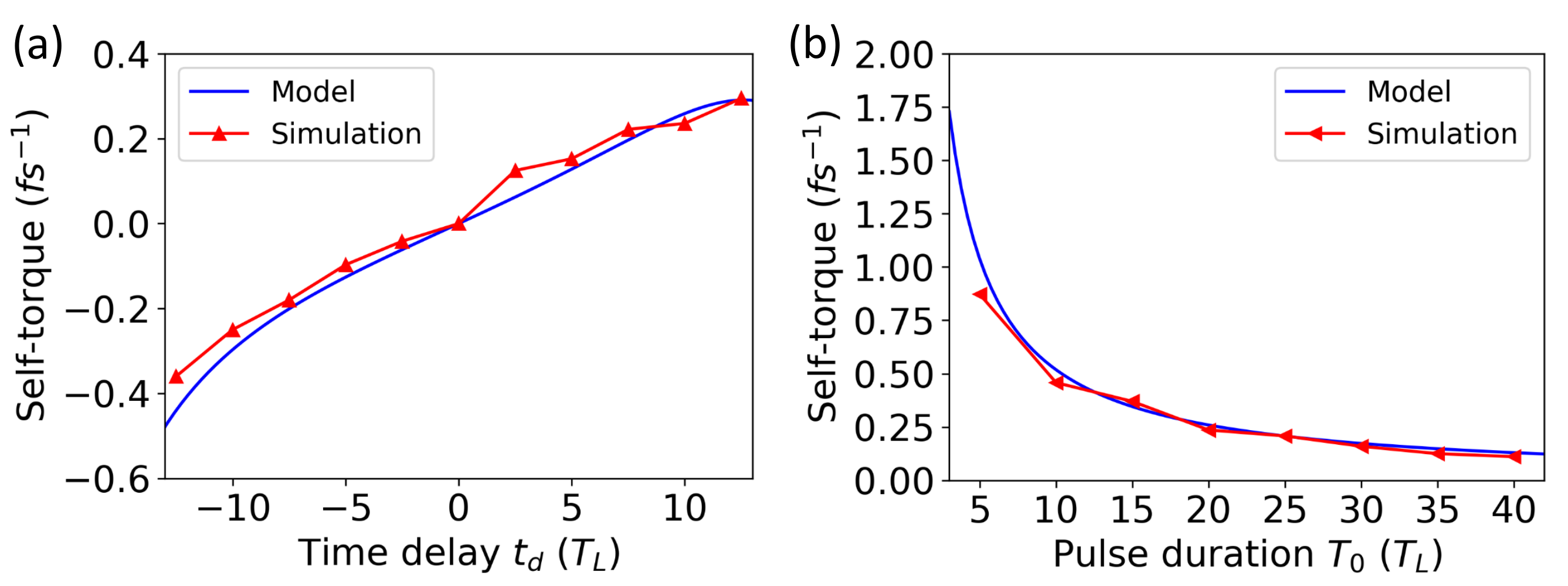}
\centering
\caption{The dependencies of the self-torque on the time delay (a) and pulse duration (b). The red triangle line is obtained from the simulations and the blue line is obtained from the theoretical model. In (a) the pulsed duration is fixed $T_0=20T_L$ and in (b) the pulse delay is set as $t_d=T_0/2$. Other simulation parameters are the same as Fig.~\ref{fig1}. }
\label{fig4}
\end{center}
\end{figure}

The self-torque can be controlled by changing the pulse duration and the delay. In Fig.~\ref{fig4} we plot the dependence of the self-torque on the pulse duration and delay. In Fig.~\ref{fig4}(a) the delay of the two pulses changes from $-12.5T_L$ to $12.5T_L$, while other parameters are the same as Fig.~\ref{fig1}. In Fig.~\ref{fig4}(b) the pulse duration $T_0$ varies from $5T_L$ to $40T_L$ and the delay is fixed as $t_d=T_0/2$. In simulations the self-torque is calculated from the frequency shift divided by the azimuthal angle offset, as shown in Fig.~\ref{fig3}(b). Both the simulations and theoretical model show that the self-torque increases with the delay $t_d$ but decreases with the pulse duration $T_0$. These can be understood from that: the OAM changing content $q(l_2-l_1)$ keeps constant for a given harmonic; therefore, the overlapping time $\tau=T_0-t_d$ determines the self-torque.

In this Letter we have presented a single example of using two laser pulses with OAM difference of 1 ($l_1=1$, $l_2=2$). In such a case the average OAM for the 5-th harmonic goes through all integer values from 5 to 10. In the case of OAM difference of $p$, for example $l_1=1$ and $l_2=p+1$, the average OAM will go from 5 to $5(p+1)$ with the spacing of $p$ according to Eq.~(\ref{eq:e4}). Another option for creating self-torqued harmonics can potentially be the usage of two pulses with same value of OAM but with different colors. We also suspect that other mechanisms of plasma harmonics generation like coherent wake emission~\cite{Quere2006} or coherent synchrotron emission~\cite{Brugge2010, Dromey2012} would be feasible as well (far-fetched conjecture is that all harmonics generation mechanisms in crystals, gases, plasmas, electron beams, etc. would yield self-torque for proper conditions). These questions can become interesting future studies.

In conclusion, we have theoretically and numerically demonstrated that intense EUV harmonics with time-varying OAM can be produced from the relativistic plasma mirror driven by two time-delayed laser pulses with different OAM values. The generated self-torqued light beam is essentially different from the fractional vortex beam. An important feature of the self-torqued harmonics is the azimuthal frequency chirp. Such an intense self-torqued EUV source may find their applications in launching selective and chiral excitation of quantum matter, imprinting OAM centrifuges and studying the transient processes sensitive to OAM.

This work was supported by the National Natural Science Foundation of China (NSFC 11674341, 11991074), the Strategic Priority Research Program of Chinese Academy of Sciences (Grant Nos. XDA25051100, XDB1603), the Chinese Academy of Sciences President's International Fellowship Initiative (No.~2018VMC0012), and the joint laser laboratory between Shanghai Institute of Optics and Fine Mechanics of Chinese Academy of Sciences and Institute of Applied Physics of Russian Academy of Sciences. The authors gratefully acknowledge the computing time granted by the National Supercomputer Center Tianhe-2 in Guangzhou and Skoltech CDISE supercomputer ``Zhores"~\cite{Zacharov2019}.

\end{document}